\documentclass[conference]{IEEEtran}
\IEEEoverridecommandlockouts
\usepackage{cite}
\usepackage{amsmath,amssymb,amsfonts}
\usepackage{algorithmic}
\usepackage{siunitx}
\usepackage{physics}
\usepackage{graphicx}
\usepackage{textcomp}
\usepackage{float}
\usepackage{caption}
\usepackage{xcolor}
\usepackage{url}
\def\BibTeX{{\rm B\kern-.05em{\sc i\kern-.025em b}\kern-.08em
    T\kern-.1667em\lower.7ex\hbox{E}\kern-.125emX}}
\begin{document}

\title{Slug-Mapper: Magnetic Scanner for \\ Ultra Low-Field MRI Scanners  
\thanks{}
}

\author{\IEEEauthorblockN{Jonathan W. Morris}
\IEEEauthorblockA{\textit{Undergrad} \\
\textit{UC Santa Cruz}\\
Santa Cruz, USA \\
jowemorr@ucsc.edu}
}

\maketitle

\begin{abstract}

Ultra-Low Field (ULF) MRI scanners offer a portable and cost-effective alternative to conventional MRI systems, but require careful calibration to compensate for reduced magnetic field strength and signal quality. In this paper, we present Slug-Mapper, a low-cost, open-source magnetic field scanner designed to map the static field $B_0$ voxel-by-voxel within the bore of a ULF MRI system. Built using a repurposed 3D printer, a Raspberry Pi, and off-the-shelf components, Slug-Mapper enables high-resolution magnetic field characterization for both passive and active shimming.
We provide a detailed overview of MRI physics, the engineering challenges of ULF systems, and the architecture and operation of Slug-Mapper. This tool bridges the gap between hardware prototyping and precise system calibration, supporting the development of accessible and portable MRI technology. 

\end{abstract}


\section{Introduction}
Magnetic Resonance Imaging (MRI) is a powerful medical imaging technique capable of producing detailed images of soft tissues, but conventional MRI scanners are expensive, difficult to maintain, and inherently non-portable \cite{marquesLowfieldMRIMR2019}. These limitations restrict access in low-resource settings and make bedside imaging impractical. Ultra-Low Field (ULF) MRI scanners \cite{liuLowcostShieldingfreeUltralowfield2021} aim to address these challenges by operating at significantly lower magnetic field strengths. While this reduction in field strength leads to lower image resolution, it enables lightweight, mobile systems that can be brought directly to patients—improving comfort, accessibility, and diagnostic speed.

However, ULF MRI scanners introduce new engineering challenges, particularly around field homogeneity and signal quality. To meet these challenges, precise mapping and calibration of the base magnetic field ($B_0$) are essential. In this paper, we provide a brief overview of MRI physics and the components involved in image acquisition. We then introduce Slug-Mapper, a low-cost, open-source device for voxel-by-voxel measurement of the $B_0$ field inside an MRI bore. This tool is designed to support both passive and active shimming by enabling 3D magnetic field mapping using affordable components and a repurposed 3D printer.

\section{MRI}
An MRI consists of three main components. A base magnetic field $B_0$, gradient coils, and Radio Frequency (RF) coils. The change in magnetization of atoms is modeled by the Bloch-Torrey PDE \cite{torreyBlochEquationsDiffusion1956}

\begin{equation}
\label{eq:BlochTorrey}
\frac{\partial M(\vec{r}, t)}{\partial t} = \left( i \gamma \Delta B(\vec{r}) - \frac{1}{T_2} \right) M(\vec{r}, t) + D \nabla^2 M(\vec{r}, t)
\end{equation}

\begin{figure}
    \centering
    \includegraphics[width=\linewidth]{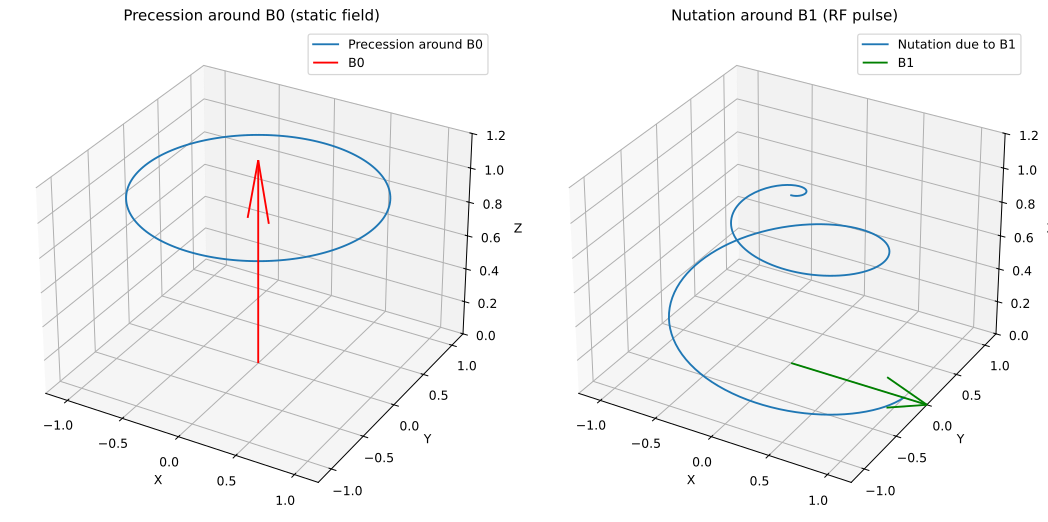}
    \caption{Illustration of how atomic spins precess around the static magnetic field $B_0$, and nutate due to the applied RF field $B_1$. Key idea is that the atom can only be measured in the transverse plane once the magnetic spin of the atom is nutated to that plane.}
    \label{fig:precession_nutation}
\end{figure}

where $\vec{r}$ is the spatial position within the MRI, $M(\vec{r}, t)$ is the magnetization at $\vec{r}$ at time $t$. $\gamma$ is the gyromagnetic ratio which is specific to the atom's quantum properties. $\Delta B(\vec{r})$ is the change in local magnetic field inhomogeneity. $T_2$ is the transverse relaxation time, which describes how quickly the spins of excited nuclei lose phase coherence in the $XY$ plane after the $B_1$ pulse nutates the spins. Figure \ref{fig:precession_nutation} illustrates the precession and nutation of the atomic spin. $D$ is the diffusion coefficient which models the diffusion of the atoms, which scales the diffusion of the atoms, and $\nabla^2$ is the Laplacian operator that in this equation, describes the net magnitude of the change in magnetization in the transverse plane

An MRI measures the instantaneous magnetization of hydrogen atoms in the water within the human body. \cite{brownMagneticResonanceImaging2014} First, a strong magnetic field called $B_0$, generated by powerful magnets, aligns the spins of hydrogen protons along the Z-axis. This establishes a baseline magnetization, which forms the foundation for the signals measured by the RF receiving (RX) coils. Gradient coils are then used to select a specific voxel in a 3-D voxel grid for spatial encoding. Next, RF transmission (TX) coils emit a RF pulse, creating a temporary magnetic field $B_1$ perpendicular to $B_0$ This causes the net magnetization vector $\vec{M}$ to nutate, or tip away from the Z-axis, rotating around the $B_1$ field into the transverse ($XY$) plane. After the RF pulse is turned off, the spins begin to precess coherently in the transverse plane, generating a rotating magnetic field. This induces a voltage in the RX coils via electromagnetic induction. The strength and decay of this signal are affected by the $T_2$ relaxation time, which varies across tissue types due to differences in water and hydrogen content — allowing MRI to clearly distinguish soft tissues. $T_1$ relaxation on the other hand refers to the time taken for the magnetization along the $Z$ dimension to return to equilibrium.

\begin{figure}
    \centering
    \includegraphics[width=\linewidth]{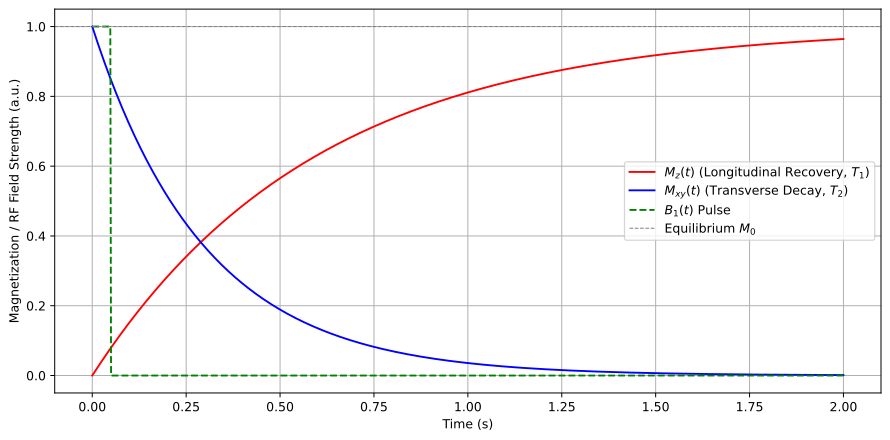}
    \caption{Graph showing the change back to equilibrium, recovery, in magnetization along the $Z$-axis ($M_{z}(t)$) and decay of magnetization in the $XY$ transverse plane ($M_{xy}(t)$). The relaxation begins immediately after $B_1$ is applied.}
    \label{fig:relaxation}
\end{figure}

\subsection{$B_0$}
The first step in building an MRI is to generate $B_0$, which is the base magnetic field inside the bore of an MRI. $B_0$ causes the atom's nuclear spins within the magnetic field to precess \cite{brownMagneticResonanceImaging2014}, or vibrate in resonance, in the direction of $B_0$. Intuitively, this is done to establish a baseline, as different tissue types return to precessing around $B_0$, after being nutated by $B_1$, at different rates due to differences in their hydrogen atom environments and concentrations. After the RF pulse is applied, the net magnetization is tipped away from alignment with the main magnetic field $B_0$ towards the transverse plane $XY$. Immediately after the application of $B_1$, the spins begin to relax back to equilibrium. This process is illustrated in Figure \ref{fig:relaxation} During this relaxation, the spins continue to precess around $B_0$, and the changing transverse magnetization induces a measurable signal in the receiver coil. By analyzing this signal, we can generate an image of the underlying atomic structure, which at a high level shows different tissues in a given area. Ultimately, uniformity and strength of $B_0$ are crucial for image quality because an irregular $B_0$ magnetic field would cause signal loss, image distortions, and variations in local relaxation behavior. \cite{winklerPracticalMethodsImproving2015}

\begin{figure}
    \centering
    \includegraphics[width=\linewidth]{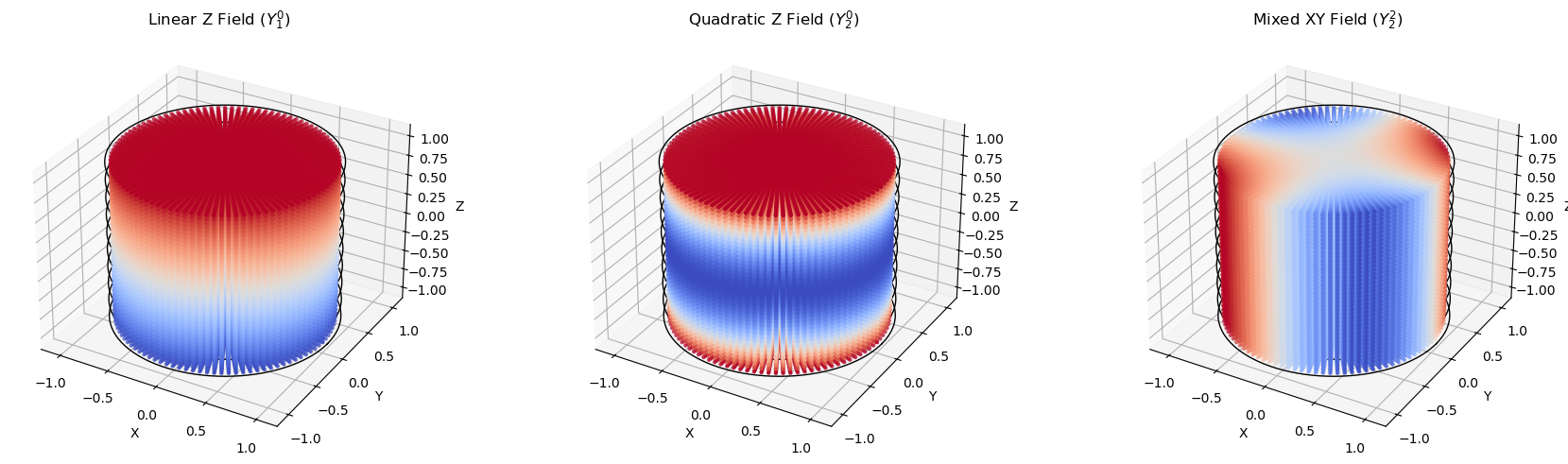}
    \caption{Illustration of how atomic spins precess around the static magnetic field $B_0$, and nutate due to the applied RF field $B_1$. Key idea is that the atom can only be measured in the transverse plane once the magnetic spin of the atom is nutated to that plane.}
    \label{fig:shimming}
\end{figure}

\subsection{Shimming}
Shimming is the process of improving homogeneity (uniformity) of the magnetic field generated by $B_0$. When shimming is performed with fixed magnets, it is referred to as passive shimming, and when shimming is performed using adjustable electromagnetic coils, it is referred to as active shimming. Passive shimming is usually only performed once during installation of an MRI \cite{ISMRM2017Shimming}, using permanent magnets made of ferromagnetic materials. Active shimming is used to adjust the electromagnetic field dynamically to correct for any inhomogeneity caused by the patient or other magnetic interference. \cite{winklerPracticalMethodsImproving2015} While in theory active shimming could replace passive shimming, both are used in practice because it is more cost-effective to let passive shimming handle the bulk of the correction. Active shimming is then used to fine-tune and compensate for residual inhomogeneities after passive shimming has been applied. Active shimming can be applied in multiple ways, linear, quadratic, or mixed, as shown by Figure \ref{fig:shimming} 

\begin{figure}
    \centering
    \includegraphics[width=\linewidth]{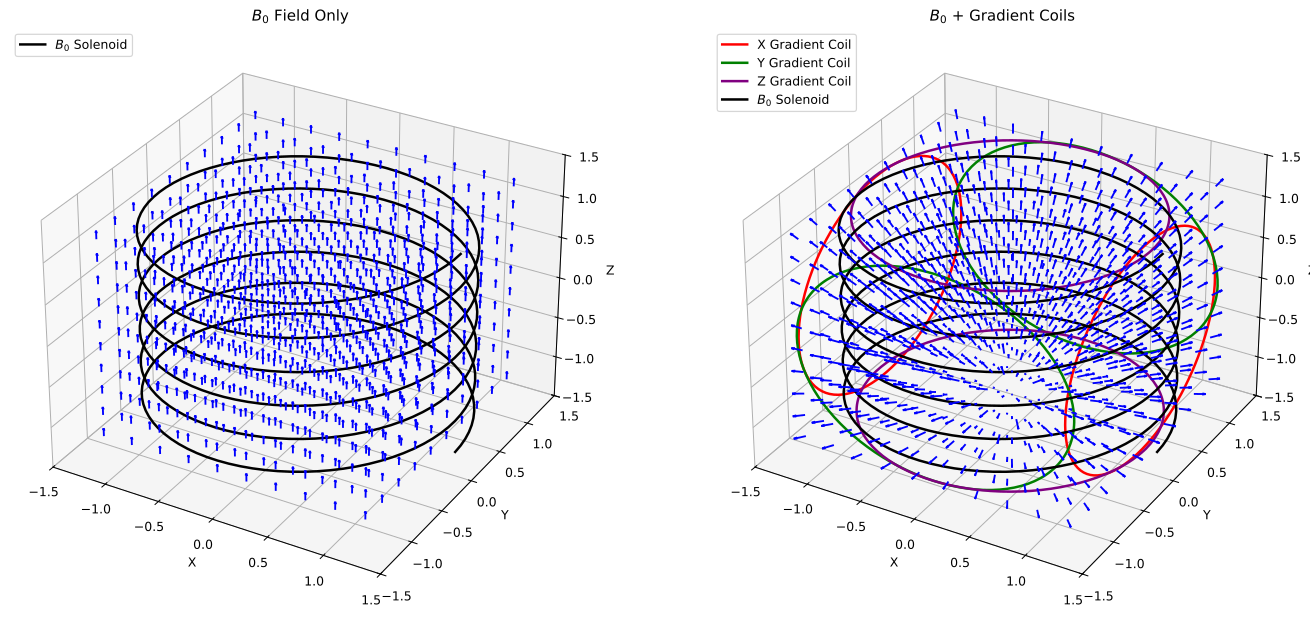}
    \caption{Effect of gradient coils on the magnetic field inside an MRI scanner. Left: The magnetic field $B_0$ generated by the main solenoid coil is homogeneous and aligned along the z-axis. Right: When gradient coils are activated, small spatial variations are introduced along the x, y, and z directions. These gradient fields are superimposed on the main $B_0$ field, creating a position-dependent magnetic field strength that enables spatial encoding of the MR signal.}
    \caption*{\textit{Note:} The magnetic field lines do not represent "flow", but instead shows the direction of the field, as described by Gauss’s law for magnetism, $\nabla \cdot \mathbf{B} = 0$, which tells us that there is no source or sink for magnetic fields.}
    \label{fig:gradient}
\end{figure}

\subsection{Gradient}
After installing the base magnets and performing shimming, a homogeneous $B_0$ field exists throughout the MRI bore. However, with a uniform magnetic field, it is impossible to determine the spatial origin of the signals received from the RF coil, as every voxel—each point in 3D space—would resonate at the same Larmor frequency \cite{larmorDynamicalTheoryElectric1897}. 

The Larmor frequency is the resonant frequency of the precession of atomic nuclei around an external magnetic field. It is defined as:

\begin{equation}
    \label{eq:larmor}
    f = \frac{\gamma}{2\pi} B_0
\end{equation}

Here, $f$ is the Larmor frequency in \si{\hertz}. $\gamma$ is the gyromagnetic ratio, expressed in radians per second per tesla (\si{\radian\per\second\per\tesla}), and is specific to the type of nucleus being measured. For MRI scanners, we are primarily concerned with the Larmor frequency of the hydrogen proton. $B_0$ is the static magnetic field strength, measured in teslas (\si{\tesla}). The significance of the Larmor frequency will be discussed further in Section~\ref{subsec:RF}.

Gradient coils superimpose small, linearly varying magnetic fields on top of the main $B_0$ field along the x, y, and z axes.\cite{brownMagneticResonanceImaging2014} This spatial variation causes hydrogen nuclei in different voxels to experience slightly different local magnetic field strengths, and thus resonate at slightly different Larmor frequencies. This frequency encoding allows us to decompose the received RF signal and isolate the contribution from each voxel. Figure~\ref{fig:gradient} shows the standard placement and orientation of the gradient coils in an MRI scanner. Gradient coils are built using copper wires to enable fast switching of current, generating magnetic fields in specific directions and magnitudes. These fields are applied at precise moments to match the pulse timing of the RF coils, enabling spatial encoding of the MR signal.

\subsection{Radio Frequency System}
\label{subsec:RF}
The final component of an MRI scanner is the Radio Frequency (RF) system, which includes both the Transmission (TX) and Receiving (RX) coils.\footnote{The RF TX and RX coils can be implemented using the same wire with a TX-RX switch in the circuit, but for simplicity, they are described as separate coils here to distinguish their functions.} \cite{gruberRFCoilsPractical2018} These coils are structurally similar to the gradient coils in that they are wire-based, but they operate at radio frequencies and serve a different purpose. Unlike gradient coils, RF coils are often made using Litz wire, which reduces losses caused by the skin effect—a phenomenon where alternating current tends to flow near the surface of a conductor at high frequencies. Litz wire increases efficiency by using many thin, individually insulated strands woven together, allowing current to distribute more evenly and minimizing skin effect losses. This design enables Litz wire to operate efficiently at radio frequencies and improves sensitivity to the weak signals induced by precessing nuclear spins.

\begin{figure}[H]
    \centering
    \includegraphics[width=\linewidth]{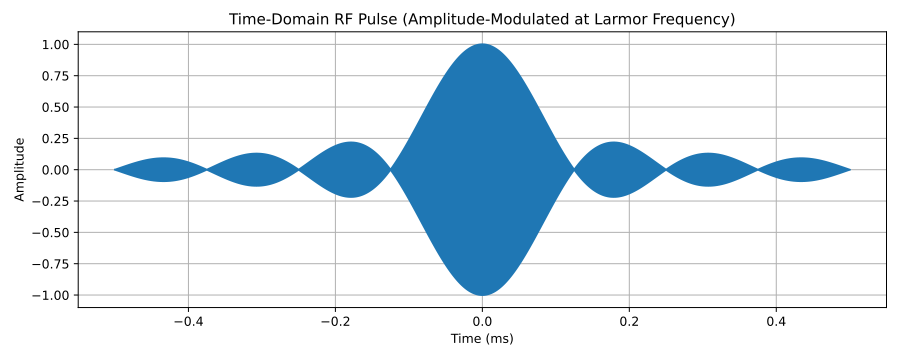}
    \caption{Example of an RF TX pulse generated by the RF coil at the Larmor frequency. The pulse follows a sinusoidal waveform modulated by an envelope, designed to match the natural precession of hydrogen nuclei. This ensures efficient nutation of the spins during slice selection.}
    \label{fig:txpulse}
\end{figure}

The RF TX coil first emits an RF pulse at the Larmor frequency, applied in the transverse plane—this is known as the $B_1$ field. The $B_1$ field causes hydrogen nuclei to nutate, tipping away from alignment with the main field $B_0$. When a gradient is applied along an axis (e.g., the z-axis), it causes the Larmor frequency to vary with position. As a result, only the spins in the slice whose Larmor frequency matches that of the RF pulse are excited. This enables spatial localization within the 3D volume. In essence, the gradient coils slightly modify the local magnetic field $B_0$, which in turn causes the Larmor frequency at each point in space to vary slightly, as described in Equation~\ref{eq:larmor}. The RF pulse can then selectively nutate hydrogen nuclei in a particular slice by matching the frequency of the pulse to the Larmor frequency at that slice.

Practically, the Larmor frequency is applied as a sinusoidal waveform over time, as shown in Figure~\ref{fig:txpulse}, to ensure that the $B_1$ field efficiently nutates the hydrogen nuclei by matching their natural precession. Imagine a swing at a playground: when a force is applied in sync with the swing’s motion, it amplifies the swing's movement; but if the force is mistimed, it can slow the swing down. Similarly, applying the RF pulse at the Larmor frequency and in phase with the spin's precession maximizes energy transfer, tipping the spin into the transverse plane for better imaging.

Right after the TX pulse is applied and the atoms in a slice are nutated, the RF RX coil begins measuring the voltage induced by the rotating transverse magnetization of hydrogen nuclei. This induced voltage is modeled using Faraday's Law:

\begin{equation}
\label{eq:faradayslaw}
\epsilon = -\dv{\Phi_B}{t}
\end{equation}

\begin{figure}
\centering
\includegraphics[width=\linewidth]{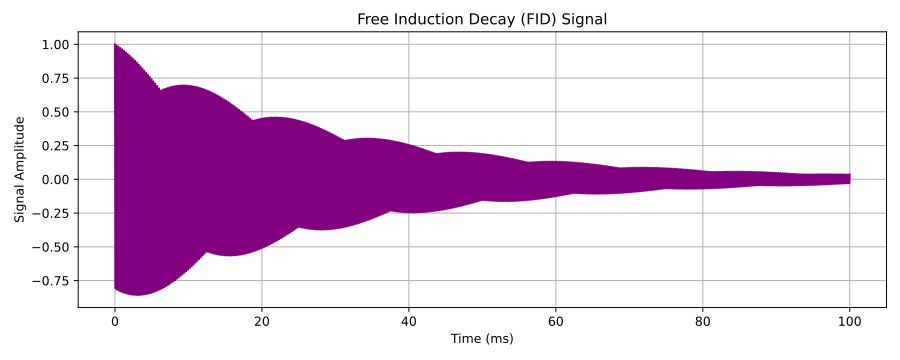}
\caption{Induced voltage measured by the RF receive (RX) coil over time, known as the Free Induction Decay (FID). The signal amplitude decreases as hydrogen nuclei gradually dephase in the transverse plane, reducing the net transverse magnetization that induces the signal.}
\label{fig:FID}
\end{figure}

where $\epsilon$ is the induced voltage, and $\Phi_B$ is the magnetic flux through the receiver coil. As nutated hydrogen nuclei precess in the transverse plane—the same plane where the TX and RX coils operate—their rotating magnetic moments induce a time-varying magnetic flux through the RX coil. This changing flux generates a measurable voltage, known as the induced signal. The gradual decay of this signal over time, due to spin dephasing, is referred to as the Free Induction Decay (FID), shown in Figure~\ref{fig:FID}. The induced voltage contains the spatial and tissue-specific information needed to form the final image.

\begin{figure}
\centering
\includegraphics[width=\linewidth]{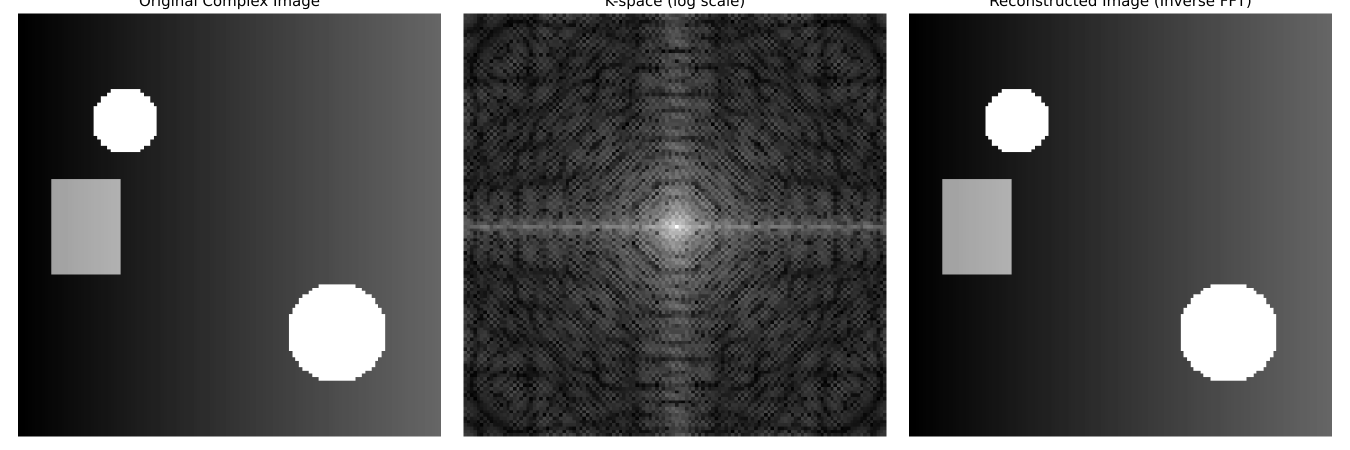}
\caption{Demonstration of how k-space encodes spatial information in MRI. The original image (left) is transformed into k-space (center), where each point stores a specific combination of spatial frequency and phase. MRI scanners fill this k-space using data received from the RF RX coil. The image is then reconstructed (right) by applying a 2D inverse Fourier transform to the k-space data.}
\label{fig:k-space}
\end{figure}

In more detail: first, the TX pulse is applied alongside a Z-gradient to nutate only the hydrogen nuclei within a specific slice of the body. Next, a Y-gradient is briefly applied to introduce a phase shift that varies across the slice—this is called phase encoding. Afterward, an X-gradient is applied during signal readout to cause the precessing spins to emit signals at position-dependent frequencies—this is frequency encoding. At this point, each voxel contributes a unique combination of phase and frequency to the FID signal. The intermediate representation of this data is known as k-space, shown in Figure~\ref{fig:k-space}, a 2D matrix of frequency and phase information.

This process is repeated multiple times, each with a different phase-encoding gradient, to fill the full k-space. Once complete, a 2D inverse Fourier transform is applied to reconstruct the image. In the final image, each pixel represents a spatial location within the slice, and the brightness (intensity) corresponds to the net transverse magnetization at that voxel.

\section{Ultra Low Field MRI}
Ultra-Low Field (ULF) MRI scanners are defined as systems with a $B_0$ field strength below \SI{0.1}{\tesla}. \cite{liuLowcostShieldingfreeUltralowfield2021} The primary goal of ULF MRI is to scale down the core imaging technology so the scanner can be brought to the patient, rather than requiring the patient to be transported to a conventional MRI suite. This makes ULF particularly valuable for critically ill patients, as the lower field strength minimizes interference with medical equipment and allows life-support lines to be routed safely. To successfully miniaturize the system while preserving image quality, researchers continue to explore alternative methods of generating the $B_0$ field and develop advanced image reconstruction techniques to compensate for the reduced signal-to-noise ratio (SNR) at lower fields. \cite{cooleyPortableScannerBrain2021}

Signal-to-noise ratio (SNR) is defined as follows \cite{ruttImpactFieldStrength1996}:

\begin{equation} 
    \label{eq:SNR}
    \text{SNR} = \frac{\mu_\text{signal}}{\sigma_\text{noise}} 
\end{equation}

This expression states that the SNR is the ratio of the mean signal intensity in a region of interest to the standard deviation of the background noise. The theoretical scaling of SNR is modeled as:

\begin{equation} 
    \label{eq:SNR_proportion} 
    \text{SNR} \propto \frac{B_0^{1.5} \cdot V \cdot \sqrt{N}}{T} 
\end{equation}

Where $B_0$ is the magnetic field strength, $V$ is the voxel volume, $N$ is the number of signal averages (i.e., how many times the scan is repeated for a slice), and $T$ is the system temperature. Generally, it is accepted that the SNR increases approximately as $B_0^{1.5}$, though this can vary depending on the coil design and sample properties. \cite{ruttImpactFieldStrength1996}

\subsection{Halbach Array}

\begin{figure*}
    \centering
    \includegraphics[width=\textwidth]{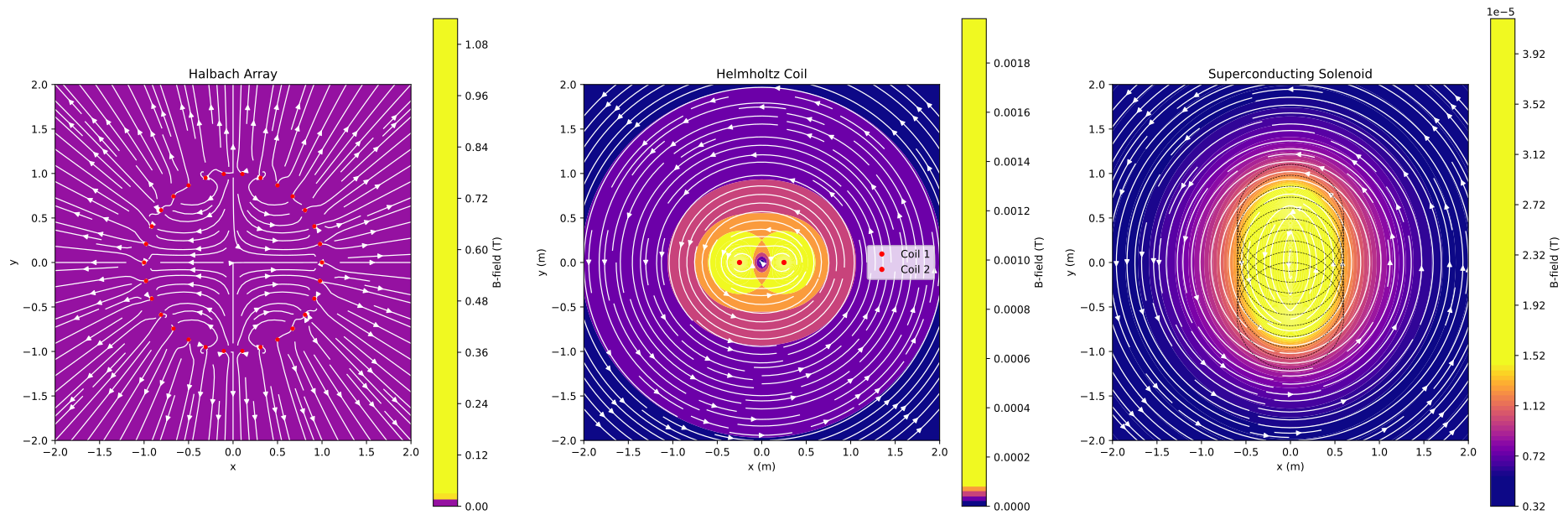}
    \caption{Magnetic field simulations of three common magnet configurations. From left to right: (1) A Halbach array, which focuses the magnetic field inside the ring of permanent magnets. (2) A Helmholtz coil, generating a near-uniform field between two current loops. (3) A superconducting solenoid, producing a strong, confined field within its cylindrical structure. The color scale represents the magnetic field strength in Tesla, with streamlines indicating the field direction. A superconducting solenoid is the most common configuration for MRIs, where the patient is laid in the bore of the solenoid.}
    \label{fig:magnets}
\end{figure*}

A Halbach array, shown in Figure \ref{fig:magnets}, is a special arrangement of permanent magnets that amplifies the magnetic field on one side of the array while canceling it on the other. \cite{cooleyDesignSparseHalbach2018} It is the most used magnet configuration for ULF MRIs, but it is not the only magnet configuration using for MRIs. In a circular configuration, this results in a homogeneous magnetic field at the center. This is the most efficient magnet configuration to get a strong field in the center of the bore. ULF MRIs use this configuration to generate a magnetic field as uniform as possible field inside the MRI. Traditional MRIs use a solenoid configuration with superconductors, however, that is not a viable option for ULF MRIs because superconductors are only magnetic at a temperature of at least $-70^{\circ}\mathrm{C}$ \cite{brownMagneticResonanceImaging2014}, exact temperature varies based on chemical composition of the superconductor. To keep the superconductors at the right temperature, liquid helium is required, which is expensive and require large machinery to move that helium through the system. This increases the cost and complexity of conventional MRIs that use a superconducting solenoid to generate $B_0$. On the other hand, Halbach arrays are commonly used in ULF MRI research as they can be formed using multiple Neodymium magnets, which are magnetic at room temperature. In Figure \ref{fig:magnets}, the Halbach array was created using just 30 such magnets. At the end of the day, any magnet configuration that can create a stable, homogeneous, field inside the MRI can be used. \cite{MathematicsPhysicsEmerging1996}

\subsection{Cooling}
Because Ultra-Low Field (ULF) MRI scanners operate with a magnetic field strength below \SI{200}{\milli\tesla}, often generated using a Halbach array of permanent magnets, they do not require liquid helium cooling. Liquid helium is typically used in conventional MRI scanners to cool superconducting magnets that produce strong $B_0$ fields. In contrast, ULF systems can use water-cooled resistive magnets or passive permanent magnets, significantly reducing size, cost, and infrastructure requirements. As a result, ULF MRI systems can be made compact enough to be transported on a cart or deployed in low-resource environments.

\subsection{Machine Learning}
Since the magnetic field generated by ULF MRIs is lower than that of conventional systems, the resulting SNR is also lower, as described by Equation~\ref{eq:SNR_proportion}. This leads to lower signal quality in the final image reconstructed from the RF signal. To mitigate this, Machine Learning (ML) algorithms are used to increase the image quality of ULF MRI scans to match quality observed from conventional MRI scanners. \cite{fesslerModelBasedImageReconstruction2010}

\section{Slug-Mapper}

\begin{figure}[H]
    \centering
    \includegraphics[width=0.48\textwidth]{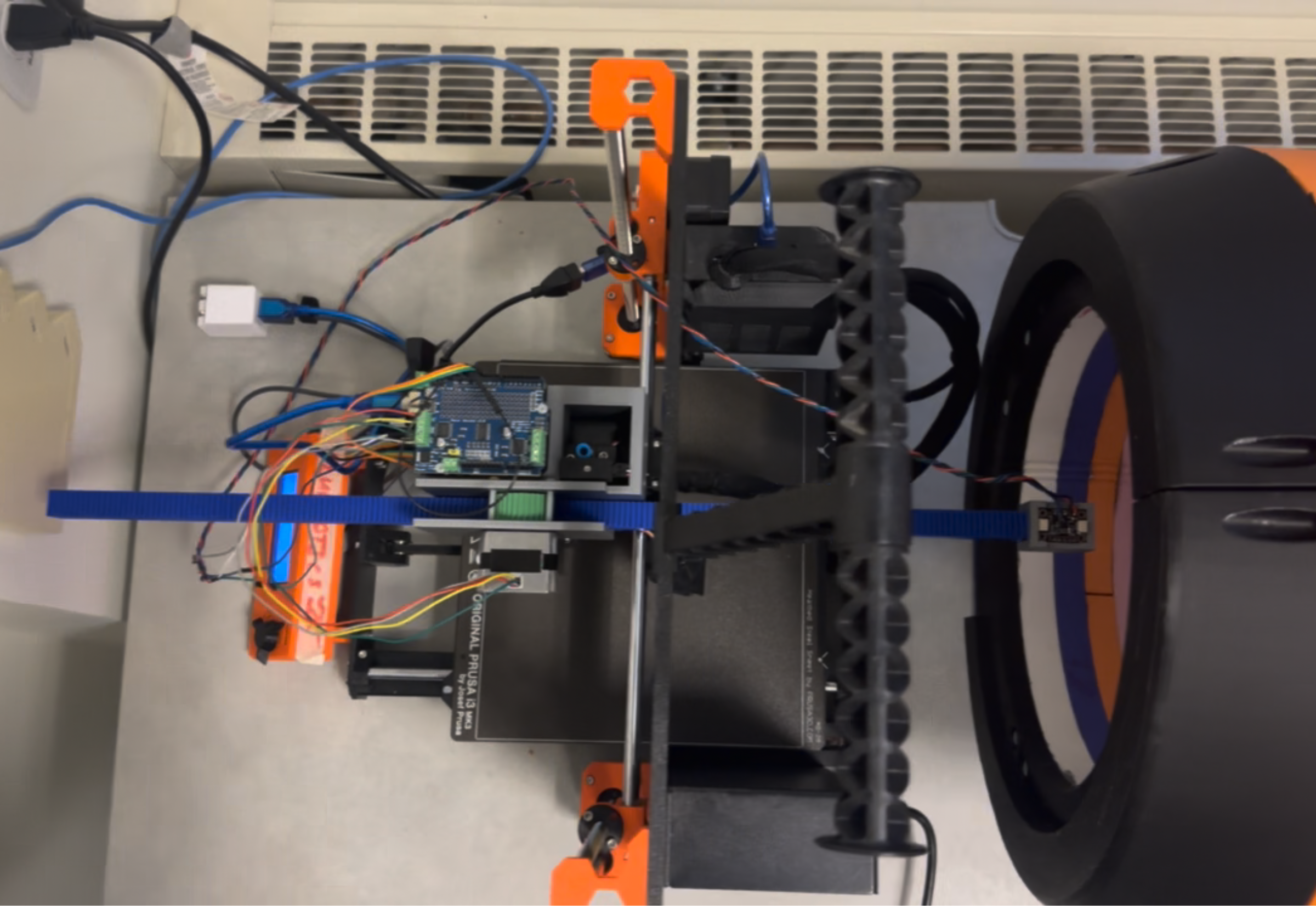}
    \caption{Image of Slug-Mapper scanning the bore of an ULF MRI.}
    \label{fig:slugmapper}
\end{figure}

Slug-Mapper is a scanner that measures the strength of the magnetic field of $B_0$ at each voxel inside the bore of an MRI, Figure \ref{fig:slugmapper}. Slug-Mapper outputs a 3D mesh with the net magnetic strength at each voxel. This information shows the inhomogeneity within the bore of the MRI, which can be corrected with passive and active shimming. A scan can also be performed to ensure that the gradient coils are creating a gradient as intended. Ultimately, Slug-Mapper allows for the calibration and testing of ULF MRIs, and it is made using ready made components purchased from Adafruit, and Digikey along with an old repurposed 3D printer.

\subsection{Components}

The Slug-Mapper system is composed of six main parts: a Prusa i3 3D printer, a Raspberry Pi Zero, an Arduino Uno with an Adafruit Motor HAT, an Adafruit NEMA 17 stepper motor, an Adafruit 3-axis magnetometer, and a custom 3D-printed rack and pinion mechanism mounted to a base frame. 

The Prusa i3 3D printer was chosen because it is already precisely calibrated in the X and Y directions, making it ideal for motion in the transverse plane of an MRI. To enable full 3D field mapping, a Z-axis was added using a rack and pinion mechanism. This mechanism is designed such that each step of the stepper motor advances the rack by exactly \SI{1}{\milli\meter}. As a result, the system achieves a spatial resolution of \SI{1}{\milli\meter} × \SI{1}{\milli\meter} × \SI{1}{\milli\meter} per voxel, allowing for fine-grained magnetic field mapping.

The Raspberry Pi Zero serves as the central controller for the system. It communicates with the Arduino Uno over SPI. The Arduino, in turn, controls the stepper motor via the Adafruit Motor HAT over I2C. The Raspberry Pi also reads data from the magnetometer over I2C, and communicates with the Prusa 3D printer via UART through a USB connection. This modular setup allows for precise, programmable control of spatial movement and field measurements, enabling voxel-by-voxel acquisition of the static magnetic field $B_0$ within the MRI bore.

\subsection{Assembly}

To assemble the Slug-Mapper, begin by flashing the Raspberry Pi Zero with the latest Raspbian Lite operating system (the non-GUI version is recommended for lower overhead). Flash the Arduino Uno with the appropriate firmware, which, along with all STL files and control code, can be found in the open-source repository.\footnote{\url{https://github.com/JonathanWMorris/slug-mapper}}

Next, attach the Adafruit Motor HAT onto the Arduino Uno and securely connect the NEMA 17 stepper motor. Ensure the motor wiring matches the board documentation—incorrect phase alignment will result in improper motor rotation or failure to step. Connect the SPI pins (MOSI, MISO, SCK, and CS) from the Arduino to the corresponding pins on the Raspberry Pi. Additionally, connect a shared ground between both boards to ensure proper logic levels.

The Adafruit 3-axis magnetometer is then connected directly to the Raspberry Pi via the I\textsuperscript{2}C interface. This includes wiring SDA, SCL, \SI{3.3}{\volt} power, and ground. All pin mappings are documented in the GitHub repository.

Next, 3D print the mechanical components: the rack-and-pinion assembly for Z-axis translation, the magnetometer mounting bracket, and the adapter for attaching to the Prusa i3 extruder head. STL files for all required components are provided in the repository. Once printed, mount the rack-and-pinion mechanism onto the Prusa frame and secure the magnetometer at the scanning end of the rail. The system should be fully constrained in X and Y by the Prusa i3 and move along Z via the stepper-controlled rack.

Finally, establish USB communication between the Raspberry Pi and the Prusa i3 using a USB-B to USB-A cable. This connection is used to control X and Y positioning via G-code commands. Power the Arduino separately using a \SI{12}{\volt} \SI{2}{\ampere} external power supply connected to the Motor HAT. Note: USB power alone is insufficient to drive the stepper motor reliably—external power is essential to avoid missed steps or brownouts during motion.

Once assembled, the Slug-Mapper system is controlled by a set of Python scripts running on the Raspberry Pi, which coordinate synchronized 3D positioning and voxel-by-voxel magnetic field sampling across the MRI bore. To ensure proper functionality, enable both SPI and I\textsuperscript{2}C interfaces via the Raspberry Pi’s raspi-config utility, and install all required Python packages as listed in the requirements.txt file provided in the GitHub repository.

\subsection{Standard Operating Procedure}
To run a scan using the Slug-Mapper, begin by opening the \texttt{main.py} script on the Raspberry Pi. Within the script, adjust the scan dimensions to match the region of the MRI bore you intend to map. These dimensions correspond to the number of voxels to sample along the X, Y, and Z axes.

Before executing the script, manually position the 3D printer head at the origin point—typically the front-bottom-left corner of the scan region. This ensures that the software’s coordinate system is aligned with the physical space being measured. Additionally, make sure the printer is mounted such that its X and Y axes are parallel to the MRI bore, and the system is level. Any tilt or misalignment can introduce spatial distortion in the resulting field map.

Once the scanner is aligned and parameters are set, run the script using:

\begin{center}
\texttt{python3 main.py}
\end{center}

The script will coordinate motor movement and magnetometer readings to generate a 3D map of the $B_0$ field inside the bore.

\section{Conclusion}

This paper presented an overview of the physics behind Magnetic Resonance Imaging (MRI), with a particular focus on Ultra-Low Field (ULF) MRI systems and their associated engineering challenges. ULF MRIs promise greater accessibility, portability, and safety, but require careful calibration and noise compensation to achieve clinically useful image quality. 

To address these needs, we introduced Slug-Mapper: a low-cost, open-source scanner designed to map the magnetic field strength $B_0$ voxel-by-voxel within the bore of an MRI system. Built using readily available components and a repurposed 3D printer, Slug-Mapper provides an effective solution for characterizing field homogeneity, diagnosing gradient behavior, and validating passive and active shimming strategies in ULF environments.

By offering a practical tool for visualizing and measuring the internal magnetic landscape of an MRI scanner, Slug-Mapper helps bridge the gap between hardware prototyping and robust system calibration. As ULF MRI systems continue to evolve, tools like Slug-Mapper will play an essential role in making high-quality medical imaging more affordable, portable, and globally accessible.

Future work will explore integrating machine learning models directly into the reconstruction pipeline and expanding Slug-Mapper to support multi-axis magnetic field mapping and real-time visualization.

\bibliographystyle{IEEEtran}
\bibliography{bib.bib}

\end{document}